%% file: main.tex
\documentclass[conference]{IEEEtran}
\usepackage[noadjust]{cite}
\usepackage{multirow} 
\usepackage{algpseudocode}
\usepackage{algorithm}
\usepackage{rotating}
\usepackage{fancyvrb}
\usepackage{kantlipsum} 
\usepackage{commath}
\allowdisplaybreaks
\usepackage{mathtools}  
\usepackage{verbatim}
\usepackage{xr-hyper} 
\usepackage{enumitem}

\usepackage{epstopdf}
\usepackage{wrapfig}
\usepackage{latexsym}
\usepackage{amssymb}
\usepackage{amsthm}
\usepackage{amsfonts}
\usepackage{amsmath} 
\usepackage{graphicx}
\usepackage{latexsym}
\usepackage{booktabs}
\usepackage{support-caption} 
\usepackage{subcaption} 
\usepackage{xspace}

\usepackage[normalem]{ulem} 

\usepackage[T1]{fontenc}
\usepackage{listings}

\ifCLASSINFOpdf
\else
\fi
\newcommand{\dq}[1]{``#1''}

\usepackage[dvipsnames]{xcolor}

\newif\ifcommentson
\commentsontrue

\newif\ifextended
\newif\ifshortver

\shortvertrue

\newif\ifrevision

\begin{document}

\title{Implementation of Accurate Per-Flow Packet Loss Monitoring in Segment Routing over IPv6 Networks}

\author{\IEEEauthorblockN{
Pierpaolo Loreti\IEEEauthorrefmark{1}\IEEEauthorrefmark{2},
Andrea Mayer\IEEEauthorrefmark{1}\IEEEauthorrefmark{2},
Paolo Lungaroni\IEEEauthorrefmark{2},
Stefano Salsano\IEEEauthorrefmark{1}\IEEEauthorrefmark{2},
Rakesh Gandhi\IEEEauthorrefmark{3},
Clarence Filsfils\IEEEauthorrefmark{3}
}
\IEEEauthorblockA{
\IEEEauthorrefmark{1}University of Rome Tor Vergata,
\IEEEauthorrefmark{2}CNIT,
\IEEEauthorrefmark{3}Cisco Systems
}
\\
\vspace{4ex}
\textbf{Accepted to IEEE HPSR 2020 conference - Version 2 - April 2020}
\vspace{-6ex}
}

\maketitle

\begin{abstract}
Segment Routing over IPv6 (SRv6 in short) is a networking solution for IP backbones and datacenters, which has been recently adopted in several of large scale network deployments. The SRv6 research, standardization and implementation activities are going on at a remarkable pace. In particular, a number of Internet Drafts have been submitted related to the Performance Monitoring (PM) of flows in an SRv6 network. In this paper we discuss the proposed PM approaches, considering both data plane and control plane aspects and focusing on loss monitoring. Then we describe the implementation of a per-flow packet loss measurement (PF-PLM) solution based on the \dq{alternate marking} method. Our implementation is based on Linux kernel networking and it is open source. We describe a platform that can be used to validate the standardization proposals from a functional perspective and the implemented solution from the performance point of view. We analyze two different design choices for the implementation of PF-PLM and evaluate their impact on the maximum forwarding throughput of a software based (Linux) router. 
\end{abstract}

\begin{IEEEkeywords}
IPv6, Performance Measurement, Segment Routing, SRv6
\end{IEEEkeywords}

%
\IEEEpeerreviewmaketitle

\input{sec/1-introduction}

\input{sec/2-standardization}
\input{sec/3-proposed-solution}

\input{sec/4-linux-implem}

\input{sec/5-results}
\input{sec/9-conclusions}

\section*{Acknowledgments}
This work has received funding from the Cisco University Research Program Fund and from the EU H2020 5G-EVE project.

\bibliographystyle{IEEEtran}
\bibliography{biblio.bib}





\end{document}

%% file: sec/1-introduction.tex
\section{Introduction}

Segment Routing for IPv6 (SRv6 in short) is the instantiation of the Segment Routing (SR)  architecture \cite{rfc8402,filsfils2015segment} for the IPv6 dataplane. SR is based on loose source routing: a list of \textit{segments} can be included in the packet headers. The segments can represent both topological way-points (nodes to be crossed along the path towards the destination) and specific operations on the packet to be performed in a node. Example of such operations are: encapsulation and de-capsulation, lookup into a specific routing table, forwarding over a specified output link, Operation and Maintenance (OAM) operations like time-stamping a packet. More in general, arbitrarily complex behaviors can be associated to a segment included in a packet. In SRv6, the segments are represented by IPv6 addresses and are carried in an IPv6 \textit{Extension Header} called \textit{Segment Routing Header} (SRH) \cite{ietf-6man-segment-routing-header}. The IPv6 address representing a segment is called SID (Segment ID). According to the SRv6 \textit{Network Programming} concept \cite{id-srv6-network-prog}, the list of segments (SIDs) can be seen as a \dq{packet processing program}, whose operations will be executed in different network nodes. The SRv6 Network Programming model offers an unprecedented flexibility to address the complex needs of transport networks in different contexts like 5G or geographically distributed large scale data centers. With the SRv6 Network Programming model it is possible to support valuable services and features like layer 3 and layer 2 VPNs, Traffic Engineering, fast rerouting. A tutorial on SRv6 technology can be found in \cite{ventre2019survey}. The standardization activities for SRv6 are actively progressing in different IETF Working Groups, among which the SPRING (Source Packet Routing In NetworkinG) WG is taking a leading role. Recently, several large scale deployments in operator networks have been disclosed, as reported in \cite{matsushima-spring-srv6-deployment}.
Performance Monitoring (PM) is a fundamental function to be performed in SRv6 networks. It allows to detect issues in the QoS parameters of active flows that may require immediate actions and to collect information that can be used for the offline optimization of the network. The most important performance parameters that need to be monitored are packet delay and packet loss ratio. A number of Internet Draft are under discussion in the IETF SPRING WG related to Performance Monitoring of flows in an SRv6 network (SRv6 PM in short). These drafts rely on existing work for performance measurement in general IP and MPLS networks and extend them for the SRv6 PM case. 
In this paper, we study and discuss the proposed SRv6 PM approaches, considering both data plane and control plane aspects. The first goal of our work is to support the standardization activity by building an open source platform for validation and comparison of proposed SRv6 PM solutions. The platform should be usable also to evaluate specific design and implementation choices through testbed experiments. In the longer term the specific PM solutions that we have implemented can become an asset for SRv6 PM on Linux routers and hosts in production. To this aim we designed and describe our open source implementation of per-flow packet loss measurement (PF-PLM), based on Linux kernel networking. To the best of our knowledge no open source implementation of SRv6 PM mechanism is available. The proposed solution relies on the \dq{alternate marking} method described in RFC 8321 \cite{rfc8321} and provides an accurate estimation of flow level packet loss (it achieves single packet loss granularity). We discuss the processing load aspects of PF-PLM in Linux software routers, comparing two implementations based on different design choices. 

%% file: sec/2-standardization.tex
\section{Performance Monitoring methodologies and their standardization}
In this section we introduce the relevant standards for PM in IP and MPLS networks and then we discuss the solutions for SRv6 network proposed in the IETF standardization.


\subsection{IP and MPLS networks}
Being able to monitor the performance metrics for packet loss, one-way delay and two-way delay is fundamental for a service provider. RFC 4656 \dq{The One-Way Active Measurement Protocol (OWAMP)} \cite{rfc4656} provides capabilities for the measurement of one-way performance metrics in IP networks, like one-way packet delay and one-way packet loss. RFC 5357 \dq{Two-Way Active Measurement Protocol (TWAMP)} \cite{rfc5357} introduces the capabilities for the measurements of two-way (i.e. round-trip) metrics. These specifications describe both the \textit{test protocol}, i.e. the format of the packets that are needed to collect and carry the measurement data and the \textit{control protocol} that can be used to setup the test sessions and to retrieve the measurement data. For example OWAMP defines two protocols: \dq{OWAMP-Test is used to exchange test packets between two measurement nodes} and \dq{OWAMP-Control is used to initiate, start, and stop test sessions and to fetch their results} (quoting \cite{rfc4656}). Note that in general there can be different ways to setup a test session: the same test protocol can be re-used with different control mechanisms. 

RFC 6374 \cite{rfc6374} specifies protocol mechanisms to enable the efficient and accurate measurement of performance metrics in MPLS networks. The protocols are called LM (Loss Measurement) and DM (Delay Measurement). We will refer to this solution as MPLS-PLDM (Packet Loss and Delay Measurements). In addition to loss and delay, MPLS-PLDM also considers how to measure throughput and delay variation with the LM and DM protocols. Differently from OWAMP/TWAMP, RFC 6374 does not rely on IP and TCP and its protocols are streamlined for hardware processing. While OWAMP and TWAMP support the timestamp format of the Network Time Protocol (NTP) \cite{rfc5905},  MPLS-PLDM adds support for the timestamp format used in the IEEE 1588 Precision Time Protocol (PTP) \cite{IEEE1588}. There are several types of channels in MPLS networks over which loss and delay measurement may be conducted. Normally, PLDM query and response messages are sent over the MPLS Generic Associated Channel (G-ACh), which is described in detail in RFC 5586. 
RFC 7876 \cite{rfc7876} complements the RFC 6374, by describing how to send the the PLDM response messages back to the querier node over UDP/IP instead of using the MPLS Generic Associated Channel. 

\begin{figure}[t!]
    \centering
    \includegraphics[width=0.48\textwidth]{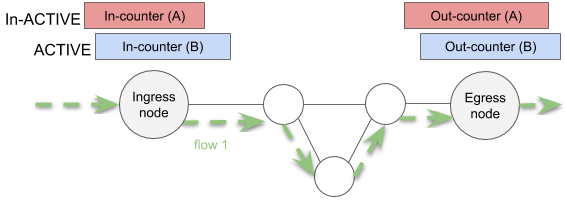}
    \caption{Alternate coloring method (RFC 8321)}
    \label{fig:alternate-coloring}
\end{figure}

Let us now focus on the procedures to monitor the packet loss experienced by a flow, from an ingress node to an egress node. In general, it is necessary to count the number of packets belonging to the flow that are sent by the ingress node and the number of packets that are received by the egress node in a reference period and then make the difference between the two counters. If we want to achieve the granularity to accurately detect single packet loss events, while the flow is active, we need to properly consider the \dq{in flight} packets, e.g. the packets counted by the ingress node but not by the egress one. The presence of in flight packets makes it difficult to obtain an accurate evaluation the number of lost packet, as discussed in RFC 8321 \cite{rfc8321}. The solution proposed by RFC 8321 is called \textit{alternate marking} method. It consists in coloring (marking) the packets of the flows to be monitored with at least two different colors, with the colors that alternate over time. For example, a continuous block of packets of a flow is colored with color A (e.g. for a configurable duration T), then the following block of packets is colored with color B (again for a duration T), and so on. Separate counters are needed in the ingress node and in the egress node to count the flow packets colored with color A and with color B. In the time interval in which the packets are colored with color B, it is possible to read counters for color A from the ingress and egress nodes, and evaluate their difference which exactly corresponds to the number of lost packet in the previous interval (see Fig.~\ref{fig:alternate-coloring}). RFC 8321 describe a generic method that can be applied to different networks. When the alternate marking method is applied to a specific network, the mechanism to color the packets must be specified. The motivations for choosing the alternate marking method and its advantages are further described in the introduction section of RFC 8321 \cite{rfc8321}.

\subsection{SRv6 networks}
The standardization activity regarding the Performance Monitoring of SRv6 networks is very active. Two internet drafts have been proposed that extend the two families of Performance Measurement solutions discussed in the previous section (OWAMP/TWAMP and MPLS-PLDM):
\begin{itemize}

\item Performance Measurement Using TWAMP Light for Segment Routing Networks \cite{gandhi-spring-twamp-srpm-05}

\item Performance Measurement Using UDP Path for Segment Routing Networks \cite{gandhi-spring-rfc6374-srpm-udp-03}
\end{itemize}

\cite{gandhi-spring-twamp-srpm-05} proposes a solution based on the extension of the TWAMP Light protocol defined in RFC 5357 Appendix I and its simplified extension Simple Two-way Active Measurement Protocol (STAMP), proposed for standardization in \cite{ietf-ippm-stamp-10}. 

\cite{gandhi-spring-rfc6374-srpm-udp-03} aims at extending and reusing the MPLS-PLDM work defined in RFC 6374 \cite{rfc6374} and RFC 7876 \cite{rfc7876}. 

Both solutions provide the possibility of measuring delay and loss of a single SRv6 flow, characterized by a SID list. The data collection takes place with test UDP packets transmitted on the same measured path. The test UDP packets collects the one way or two way PM data and makes them available to the node which requested the measurement.

The solution in \cite{gandhi-spring-rfc6374-srpm-udp-03} is likely a bit more mature (its standardization started in March 2018) and it includes more features with respect to the TWAMP based one. For example it includes a Return Path TLV used to carry reverse SR path information. It also defines a packet to collect counters for loss measurement and timestamps for delay measurement with a single message.

%




%% file: sec/3-proposed-solution.tex

\section{SRV6 Loss Measurement}

\begin{figure}[t!]
    \centering
    \includegraphics[width=0.48\textwidth]{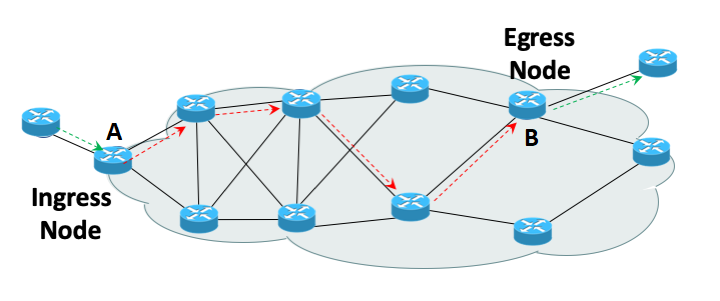}
    \caption{Reference SRv6 network scenario for Performance Monitoring}
    \label{fig:pm-data-plane}
\end{figure}

In this section we detail a Loss Measurement (LM) solution compliant with the two proposed drafts (\cite{gandhi-spring-twamp-srpm-05} and \cite{gandhi-spring-rfc6374-srpm-udp-03}). 
Our reference network scenario is depicted in Figure \ref{fig:pm-data-plane}. We have an SRv6 network domain. IP traffic arrives at an ingress edge node of the network where it can be classified and encapsulated in an SRv6 flow. In SRv6 terminology, an \textit{SR policy} is applied to the incoming packets. The SR policy corresponds to a \textit{SID List} that is included in the Segment Routing Header (SRH) of the outer IPv6 packet. The outer IPv6 packet crosses the network (according to its SID list) and arrives to the egress edge node where the inner IP packet is decapsulated (the outer IPv6 and SRH are removed). For example in the figure \ref{fig:pm-data-plane}, node A acts as ingress node while node B as egress node for the green packets. The ingress node A applies the SR policy, i.e. it writes the \textit{SID list} into the SRH header.

\subsection{Packet Counting}

In order to perform Per-Flow Loss Measurement we need to implement packet counters associated to SRv6 flows, both in the ingress node and in the egress node. For our purposes, an SRv6 flow corresponds to an SR policy, i.e. to a SID List. We want to be able to explicitly activate the counting process for a set of flows (identified by their SID Lists).
In an ingress node this means to process all outgoing SRv6 packets and count the packets belonging the set of monitored flows (by comparing the SID List of the outgoing packets with the SID Lists of the monitored flows). Likewise, in an egress node, this means to process all incoming SRv6 packets, check if the packets belong to the set of monitored flows and increment the counters as needed. 

These counting operations can have a high computing cost for a software router, so it is important to carefully design their implementation (see section \ref{sec:impl} to evaluate their impact on the processing performance).


\subsection{Traffic Coloring}

The internet drafts \cite{gandhi-spring-twamp-srpm-05} and \cite{gandhi-spring-rfc6374-srpm-udp-03} do not specify how the coloring must be implemented and thus several solutions are possibles. In case of SRv6, we considered two solutions: i) modification of the DS field, previously known as IP Type of Service (TOS) field; ii) encoding the color in a SID of the SID list present in the SRv6 header.

The first solution is simple but has some drawbacks: the number of bits available in the DS field is limited (6 or 8) and they are considered precious. Using two colors we need a bit, in addition we can use a second bit to differentiate between colored traffic to be monitored and uncolored traffic not to be monitored. This can be useful to avoid comparing the full SID List to decide if a packet is part of a flow under monitoring or not. In our current implementation described in section~\ref{sec:impl} we have used two bits of the DS field.

The solution that encodes the color in a SID, exploits the fact that according to \cite{id-srv6-network-prog} an IPv6 address representing a SID is divided in LOCATOR:FUNC:ARGS. The LOCATOR part is routable and allows to forward the packet towards the node where the SID needs to be executed. The FUNC and ARG parts are processed in the node that executes the SID. In particular, the ARG part can reserve a number of bits for the alternate marking procedures. This may allow using more than two colors. This solution however has an implementation drawback: due to the variable position of these bits, implementing an hardware processing solution is much harder and can be out of reach for current chips that need to operate at line speed. Moreover periodically changing the ARG bits in a SID of a running flow can cause an interference with the SRv6 forwarding plane (e.g. for Equal Cost MultiPath) when the SID is used as IPv6 destination addresses.

\subsection{Data Collection}


The two internet drafts proposed for the PM of SRv6 specify the use of dedicated protocols for the collection of meters and the loss evaluation.
Both standards are based on the sending of a UDP packet (query) by the ingress node (called Sender in \cite{gandhi-spring-twamp-srpm-05}) to the egress node (called Reflector in \cite{gandhi-spring-twamp-srpm-05}). The packet is used to collect the counters of a given color, in order to evaluate the loss. If path monitoring is bidirectional, the Reflector sends to the Sender a response packet that goes through the network in the reverse direction, collecting the counters of the return path. 
The draft \cite{gandhi-spring-twamp-srpm-05} specifies that the query and the response packets must comply with the TWAMP light or STAMP protocol format. Both packets use a fixed structure.

The UDP query packet includes the following fields:
\begin{itemize}
\item  Sender Sequence Number: a 32 bit number incremented by one each query message;
\item  Block Number:  The color of the direct path;
\item  Sender Counter: the output counter of the Sender.
\end{itemize}

The UDP response packet includes the following fields:
\begin{itemize}
\item  Receiver Sequence Number: a 32 bit number incremented by one each response message;
\item  Receiver Counter: the input counter of the direct path;
\item  Transmit Counter: the output counter of the Reflector;
\item  Block Number: the color of the reverse path;
\item  the three fields of the Query Packet.
\end{itemize}










%% file: sec/4-linux-implem.tex
\section{Linux Implementation}\label{sec:impl}
The Per Flow Packet Loss Monitoring (PF-PLM) system has been realized in Linux extending the kernel based SRv6 implementation and different frameworks for packet processing, namely Netfilter/Xtables and \textit{IP set}. In particular, we will present and compare a simpler solution based on \textit{iptables} and a more efficient one based on \textit{IP set}. All the software components we have developed are available as open source \cite{netgroup-srv6-pm}. In the following subsections we will provide some basic tutorials on the tools that we have extended and the descriptions of our contributions. 


\subsection{Linux SRv6 subsystem}

The Linux kernel SRv6 subsystem \cite{lebrun2017implementing} supports the basic SRv6 operation described in \cite{ietf-6man-segment-routing-header} and most of the operations defined in \cite{id-sr-service-programming}. A Linux node can classify incoming packets and apply SRv6 policies, e.g. encapsulate the packets with an outer packet carrying the list of SRv6 segments (SIDs). A Linux node can associate a SID to one of the supported operations, so that the operation will be executed on the received packets that have such SID as IPv6 Destination Address. More details on the Linux SRv6 implementation with a list of the currently supported operations can be found in \cite{ahmedperformancefull}.

\subsection{Linux Netfilter/Xtables/iptables subsystem} 

The kernel-space Netfilter/Xtables allows the system administrator to insert chains of rules for the processing of packets inside predefined tables (raw, mangle, filter, nat). Each table is associated with different types of packet processing operations. In each chain, packets are processed by sequentially evaluating the rules in the chains. As shown in Fig.~\ref{fig:overall-pm-net-stack}, there are 5 processing phases (PREROUTING, INPUT, FORWARD, OUTPUT, POSTROUTING) in which the default chains associated with specific tables are processed. Moreover, it is possible to create additional chains as needed. For example, in the POSTROUTING phase, the default postrouting chains associated to the mangle and nat table are processed. \textit{Iptables} is a user-space CLI (Command Line Interface) utility program that allows a system administrator to configure the tables/chains/rules provided by the Netfilter/Xtables subsystem. Fig.~\ref{fig:overall-pm-net-stack} shows two additional chains (BLUE-CHAIN, RED-CHAIN) that we added in our packet loss monitoring solution, visible in the rightmost part of the figure. 
For simplicity we will use \textit{iptables} to refer in general to the Netfilter/Xtables/iptables subsystem, including the IPv6 specific modules.

\textit{Iptables} is highly modular and can be extended. In particular, it is possible to develop a custom packet matching module, to specify a rule which refers to the custom module name and includes extra commands that depend on the specific extension. We have followed this approach, as described in \ref{SRH-match-module}.

\subsection{\textit{Iptables} based PF-PLM implementation}
\label{sec:iptables-implementation}

In order to evaluate Per-Flow Packet Loss metrics, we need to collect information on both Ingress and Egress nodes in a coordinated way.  



\begin{figure}[h!tb]
    \centering
    \includegraphics[width=0.48\textwidth]{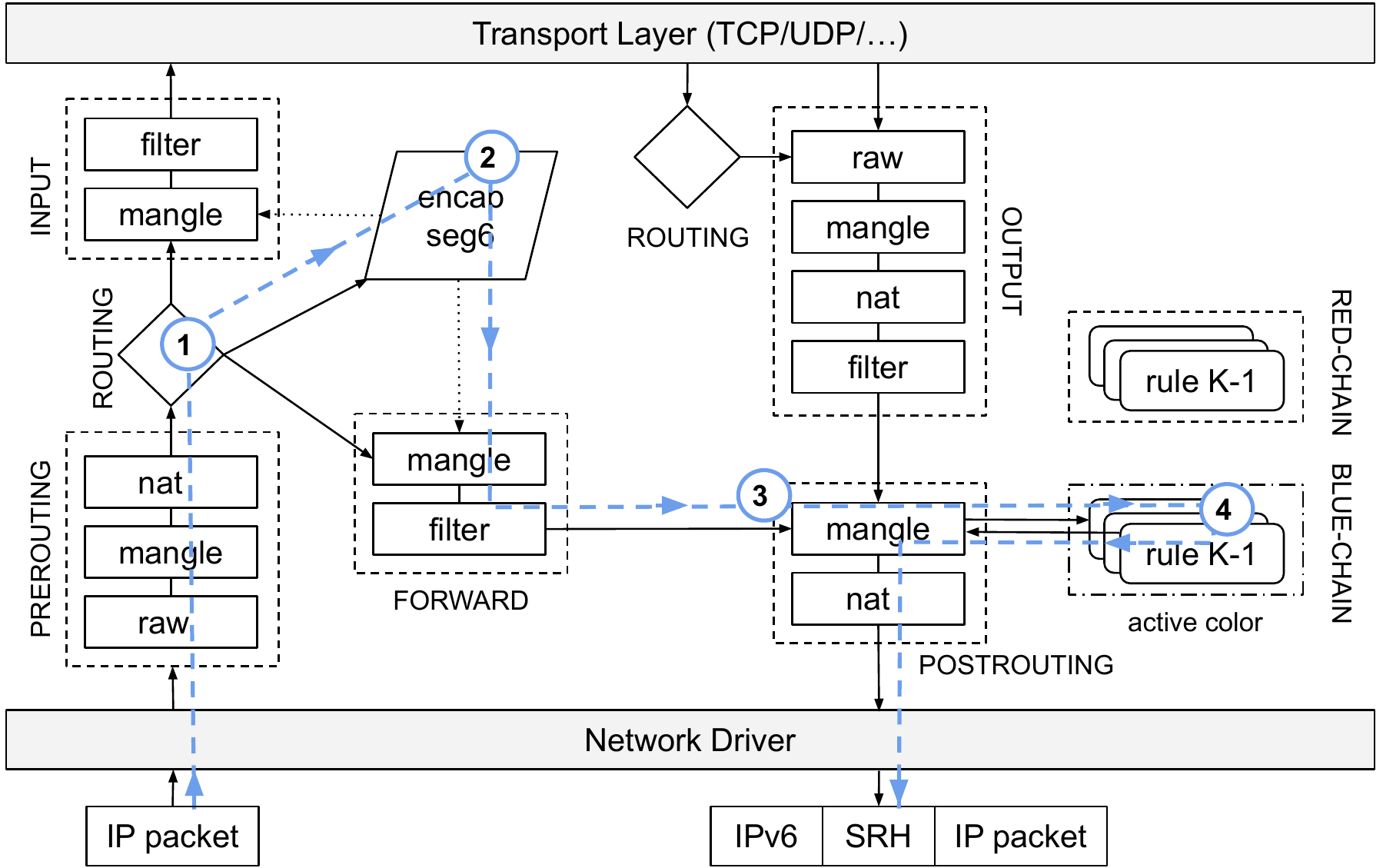}
    \caption{Packet processing in the Ingress node}
    \label{fig:overall-pm-net-stack}
\end{figure}

In the ingress node, the traffic is classified (based on IP Destination Addresses) and can be associated to an SRv6 policy (step 1 in Fig.~\ref{fig:overall-pm-net-stack}). This means that a matching packet can be encapsulated in an outer IPv6 packet with a new IPv6 header followed by the Segment Routing Header with the proper Segment List (SID List) (step 2 in Fig.~\ref{fig:overall-pm-net-stack}). The encapsulated packet continues its journey in the networking stack until he reaches the POSTROUTING chain of the mangle table (step 3 in Fig.~\ref{fig:overall-pm-net-stack}). In this chain, we put a \dq{jump} rule with the aim to divert all SRv6 traffic to a custom chain for statistic collection purposes (step 4 in Fig.~\ref{fig:overall-pm-net-stack}). In particular, we have two custom chains, referred as \textit{coloring chains}, as we need to use two colors for traffic marking according to the \dq{alternate marking} approach. Each chain contains a set of rules, each rule is used to match on a specific SID list included in the SRH header. At any given time, only one of the two chains is \textit{active} and is referred to by the \dq{jump} rule. 

When an SRv6 packet enters the active \textit{coloring chain}, the rules contained in the chain are applied to the packet in sequence until:
1) There is a rule that matches the SID list of the packet. The match counter of this rule is incremented. The packets needs to be colored with the color corresponding to the active coloring chain. When the coloring based on DS field is used, the \textit{color bit} of the DS field will be set to the proper color (0 or 1). Once this is done, the packet processing comes back to the calling chain, i.e. the POSTROUTING chain in the mangle table.
2) All rules have been considered and no match with SID list of the packet has been found, so the packet processing simply jumps back to the calling POSTROUTING chain.

\begin{figure}[h!tb]
    \centering
    \includegraphics[width=0.48\textwidth]{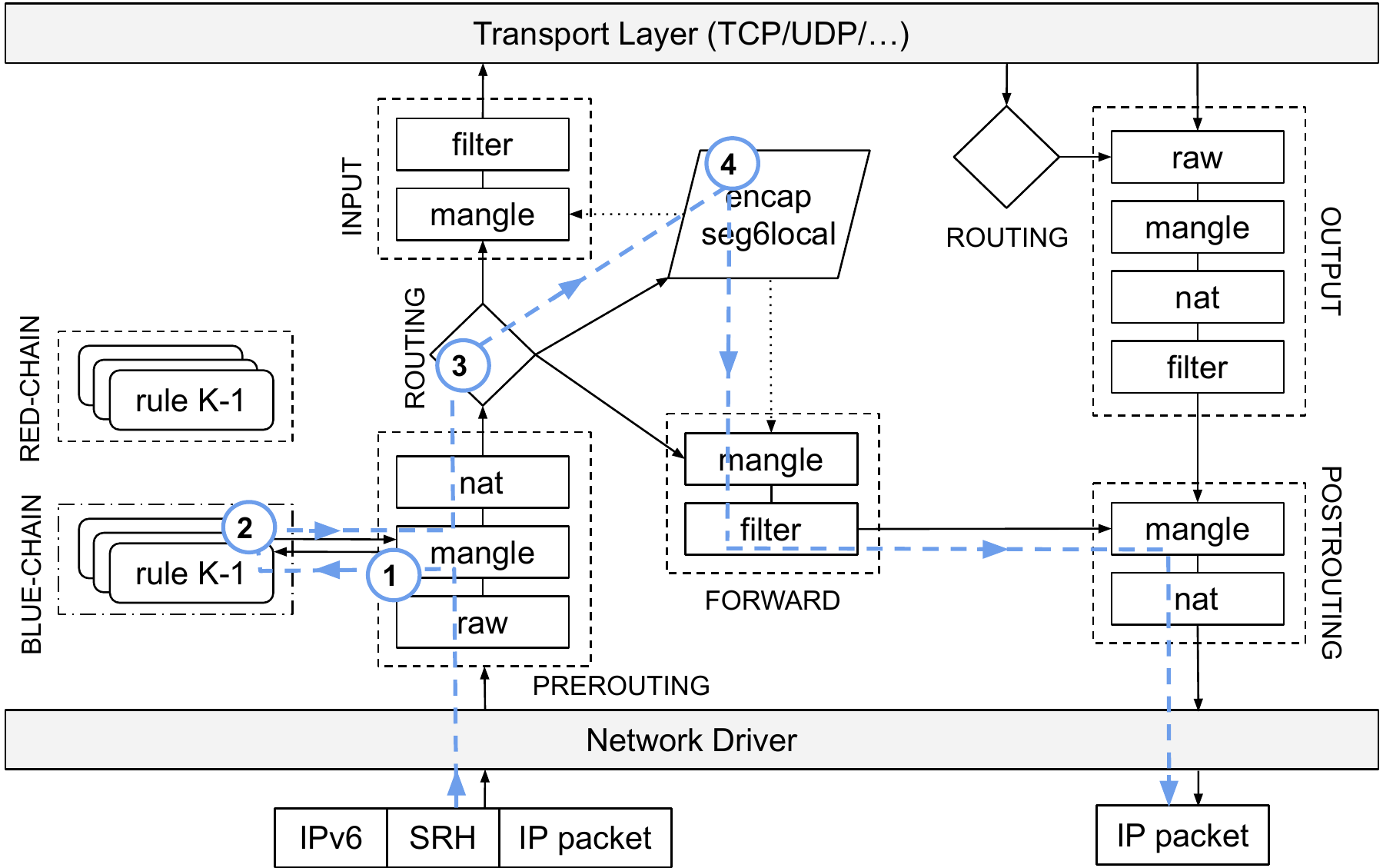}
    \caption{Packet processing in the Egress node}
    \label{fig:net-stack-egress}
\end{figure}

In the egress node, the packets of the flows to be monitored enter from an incoming interface and arrive encapsulated in an IPv6 outer header. Therefore we need to match on the SID List of incoming packet. We use the PREROUTING chain of the mangle table to match on the different colors, considering the color bit of the DS field (step 1 of Fig.~\ref{fig:net-stack-egress}). Then for each color, we process the packet in a chain which includes one rule for each SID List that we want to monitor (step 2 of Fig.~\ref{fig:net-stack-egress}). We refer to this custom chain as \textit{color-counting chain}. 
When a rule matches the SID List in the packet, the match counter of the rule is incremented. After the processing in the color-counting chain, the normal packet processing continues. As we are in the egress node, the destination address will be a SID that corresponds to a packet decapsulation operation. This SID is matched in the routing operation (step 3 of Fig.~\ref{fig:net-stack-egress}), then the decapuslation operation is executed by the SRv6 kernel processing (encap seg6local, step 4 of Fig.~\ref{fig:net-stack-egress}). 

\subsection{Improvements to the SRH match extension}
\label{SRH-match-module}

The Netfilter/Xtables already provides a module extension that matches packets based on the Segment Routing header of a packet. The module is named \texttt{ip6t\_srh} and it has been included in the Linux kernel tree since version 4.16. With the \texttt{ip6t\_srh} extension, it is possible to write Netfilter/Xtables rules that match the current SID, the next SID and the number of the left segments in a SRH. It is not possible to match packets by comparing a full SID list. Therefore, we extended \texttt{ip6t\_srh} to support matching on the SID list. 

The improvements to the \texttt{ip6t\_srh} module are needed in the kernel-space and in the user-space sides. In the kernel-space side, we have allowed to store the SID list provided by user-space associated to a rule. Furthermore, we have also created a match function that compares this SID list with the one present in the examined SRv6 packet. When the two SID lists are identical, the match function returns successfully and we proceed with the execution of further match extension modules present in the rule and/or with the execution of a target action. If the two lists differ, the match function returns an error that allows Netfilter/Xtables to move onto the next rule, if any.


For what concerns the user-space side, changes have focused on \texttt{lib6t\_srh.c}. This file is used to parse the commands provided by the user on the CLI and then to create, initialize and populate the data structures required to build a rule. The same data structures are used by \texttt{ip6t\_srh} module to carry out matches on SRv6 packets. Therefore, we have added the parsing function that reads and validates a list of SIDs supplied by the user and the fields used to store the validated SID list. We have also implemented the helper functions to retrieve and display the SID list and the number of matched packets for each rule.


\begin{lstlisting}[label={lst:iptables-us},caption={Iptables userspace tool},basicstyle=\ttfamily]
1) ip6tables -A POSTROUTING -t mangle   \
     -m rt --rt-type 4 -j blue-chain

2) ip6tables -A blue-chain -t mangle    \
    -m srh --srh-sid-list sid0,sid1     \
    -j --set-tos 0x01/0x01
\end{lstlisting}

The command (2) shown in Listing \ref{lst:iptables-us} adds a rule to the \textit{blue-chain} of the mangle table and asks Netfilter/Xtables framework to use the \textit{srh} module to match a SRv6 packet considering the SID list indicated by the attribute \texttt{-{}-srh-sid-list}. In this example, the action (\texttt{-j -{}-set-tos 0x01/0x01}) specified in the target consists in marking the packet by setting the first bit of the \textit{DS} field of the outer IPv6 header.

\subsection{Performance issues of \textit{iptables} based PF-PLM}

The rules that are included in the coloring chains (ingress node) and in the color-counting chains (egress node) are tested sequentially, until a matching rule is found (or all the rules have been tested). 


This sequential scan approach can have a significant impact on the node throughput due to the processing load for comparing the SID list in the SRH packets with the SID lists in the coloring chain or in the color-counting chain. On average, the processing load increases linearly with the number of rules. This number of rules corresponds to the number of flows for which we want to monitor the packet loss. In the performance experiments we will analyze the impact of the number of monitored flows on the node throughput due to the sequential scan approach.



\subsection{The \textit{IP set} framework} \label{ipset-framework}

\textit{IP set} \cite{ipset} is an extension to Netfilter/Xtables/iptables that is available through Xtables-addons \cite{xtables-addons}. \textit{IP set} allows to create rules that match entire \textit{sets} of elements at once. Unlike normal \textit{iptables} chains, which are stored and traversed linearly, elements in the sets are stored in indexed data structures, making lookups very efficient, even when dealing with large sets.
Depending on the type, an IP set may store IP host addresses, IP network addresses, TCP/UDP port numbers, MAC addresses, interface names or combination of them in a way, which ensures lightning speed when matching an entry against a set.
To use \textit{IP set}, we create and populate uniquely named sets using the \textit{IP set} command-line and then we can reference those sets in the match specification of one or more \textit{iptables} rules. 

By using \textit{IP set}, the \textit{iptables} command refers the set with the match specification \texttt{-m set -{}-set foo dst}, which means \dq{match packets whose destination is contained in the set with the name \textit{foo}}.
As a result, a single \textit{iptables} command is required regardless of the number of elements in the \textit{foo} set. If we want to achieve the same result with the use of \textit{iptables} only, it would be necessary to create a chain and insert as many rules as the elements contained in the \textit{foo} set.
\textit{IP set} provides different types of data structures to store the elements (addresses, networks, etc). Each set type has its own rules for the type, range and distribution of values it can contain. Different set types also use different types of indexes and are optimized for different scenarios. The best/most efficient set type depends on the situation. 
The hash sets offer excellent performance in terms of speed of the execution time of lookup/match operation and they fit perfectly to our needs.
Assuming to insert $N$ IP addresses in the hash set \textit{foo}, the cost of searching for an address is asymptotically equal to $O(1)$. Conversely, the same operation with \textit{iptables} would have a cost of $O(N)$.

The \textit{IP set} framework is designed to be extensible: there are several header files to be included that contain the helper functions and templates which, through an intelligent use of C macros and callbacks, allow to redefine and adapt the code to our needs. 

\subsection{The \textit{IP set} based PF-PLM implementation}

Using the \textit{iptables} PF-PLM implementation described in \ref{sec:iptables-implementation}, the packet processing throughput decreases when the number of flows to be monitored increases. 
To overcome this shortcoming, we have designed and implemented an enhanced solution based on \textit{IP set}. 


The implementation of \textit{IP set} does not natively allow to store elements of \texttt{SID list} type within a hash set. Therefore, in order to use \textit{IP set} for our purposes we have patched it by creating a new hash set called \texttt{sr6hash} so that we can insert the SID lists that we want to monitor.

To support the new \texttt{sr6hash} hash type, we patched the \textit{IP set} framework on both the user-space and the kernel-space sides. In the user-space side, we defined a new data structure, the \texttt{nf\_srh}, which contains the SRH header with a SID list whose maximum length is fixed and set at compilation time (16 SIDs in our experiments). Moreover, two new functions have been added to the \textit{IP set} framework libraries: the parse function (\texttt{ipset\_parse\_srh()}) and the print function (\texttt{ipset\_print\_srh()}).
The \texttt{ipset\_parse\_srh()} is used to parse the SID list supplied by the user with the \textit{IP set} CLI and to create and populate the \texttt{nf\_srh} data structures. The \texttt{ipset\_print\_srh()} is used to display all the SID lists in a given \texttt{sr6hash} set along with their match counters.


\begin{lstlisting}[label={lst:ipset-us},caption={IPset userspace tool.},basicstyle=\ttfamily]
1) ipset -N blue-ht sr6hash counters

2) ipset -A blue-ht                     \ 
     2001:db8::1,2001::db8::2

3) ip6tables -A blue-chain -t mangle    \
     -m set --match-set blue-ht         \
     -j --set-tos 0x01/0x01
\end{lstlisting}

In listing \ref{lst:ipset-us} we show the patched user-space commands used to: 1) create a hash set of type \texttt{sr6hash}; 2) add a SID list consisting of two SIDs; 3) add an \textit{iptables} rule, in which the match is performed on the hash set \textit{blue-ht} and the action consists in marking the first bit of the \text{DS} (\text{tos}) field in the outer IPv6 packet.

In the kernel part, we introduced a new \textit{IP set} module which is the implementation of the hash set \texttt{sr6hash}. In particular, we have defined:


\begin{itemize}

\item the data structure of the element of the hash set (\texttt{hash\_sr6\_elem}) which contains the SRH with SID list to be stored. The user-space and kernel-space data structures are identical so that it facilitates the exchange of information between the two contexts;

\item the equality function \texttt{hash\_sr6\_data\_equal()} to compare two SID lists;

\item the functions \texttt{hash\_sr6\_kadt()} and \texttt{hash\_sr6\_uadt()} which are used for adding, deleting an element to/from the hash set and testing the membership.

\item the \texttt{ip\_set\_type} structure where we set the properties, policies and extensions supported by our module. Therefore, this structure is used as a "glue" that sticks parts together and is used to, actually, register the hash set when the module is loaded into the kernel and to deregister it when it is unloaded.

\end{itemize}



\subsection{Data collection}

To test the effectiveness of the proposed solution we developed a prototype implementation of the Sender and Reflector based on the python Scapy project \cite{scapy} that support SRv6 packets. Both the Sender and the Reflector periodically change the active color. The Sender reads the local counter and generate the query packet that is sent using the SRv6 path. The Reflector receives the packet, reads the missing counters and sends the response packet back to the Sender, that eventually is able to evaluate the packet loss. An open-source implementation of the python code is available online \cite{netgroup-srv6-pm}.

%% file: sec/5-results.tex
\section{Results}

\begin{figure}[t!]
    \centering
    \includegraphics[width=0.48\textwidth]{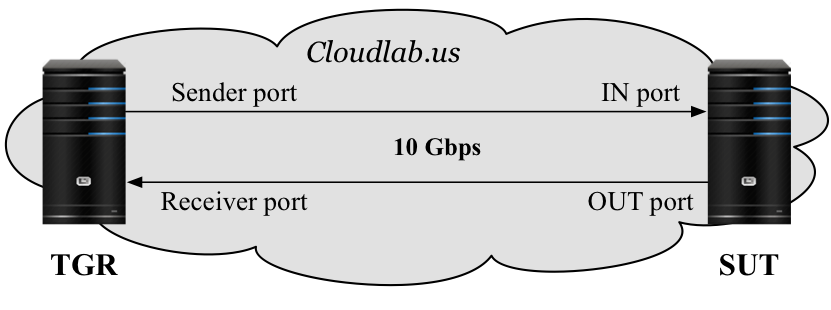}
    \caption{Testbed architecture}
    \label{fig:testbed}
    \vspace{-3ex}
\end{figure}

\subsection{Testbed Description} \label{sec:sr_test}

Figure~\ref{fig:testbed} depicts the testbed architecture, made of two nodes denoted as \textit{Traffic Generator and Receiver (TGR)} and \textit{System Under Test (SUT)}.  
In our experiments we only consider the traffic in one direction: the packets are generated by the TGR on the Sender port, enter the SUT from the IN port, exit the SUT from the OUT port and then they are received back by the TGR on the Receiver port. Thus, the \textit{TGR} can evaluate all different kinds of statistics on the transmitted traffic including packet loss, delay, etc. 

The testbed is deployed on the CloudLab facilities \cite{ricci2014introducing}, a flexible infrastructure dedicated to scientific research on the future of Cloud Computing. Both the TGR and the SUT are bare metal servers with Intel Xeon E5-2630 v3 processor with 16 cores (hyper-threaded) clocked at 2.40GHz, 128 GB of RAM and two Intel 82599ES 10-Gigabit network interface cards. The SUT node runs a vanilla version of Linux kernel 5.4 and hosts the PF-PLM system. We consider two configurations to evaluate the SUT performance:
\begin{enumerate}
\item the SUT is configured as the ingress node of the SRv6 network, i.e. it executes the encapsulation operation.
\item the SUT is configured as the egress node of the SRv6 network, i.e. it executes the decapsulation operation.
\end{enumerate}

In the TGR node we exploit TRex \cite{trex-cisco} that is an open source traffic generator powered by DPDK \cite{dpdk}. 
We used SRPerf \cite{ahmedperformance}, a performance evaluation framework for software and hardware implementations of SRv6, which automatically controls the TRex generator in order to evaluate the maximum throughput that can be processed by the SUT. The maximum throughput is defined as the maximum packet rate at which the packet drop ratio is smaller then or equal to 0.5\%. This is also referred to as Partial Drop Rate at a 0.5\% drop ratio (in short PDR@0.5\%). 
Further details on PDR and insights about nodes configurations for the correct execution of the experiments can be found in \cite{ahmedperformance}.

\subsection{SUT Performance} \label{sec:perf}

To carry out the performance experiments on the SUT, we crafted IPv6 UDP packets encapsulated in outer IPv6 packets (78 byte including all headers). The outer packets have an SRH with a SID list of one SID. We repeated each test four times (note that, as described in  \cite{ahmedperformance}, each test includes a large number of experiments and repetitions to estimate the maximum throughput).

\begin{figure}[t!]
    \centering
    \includegraphics[width=0.48\textwidth]{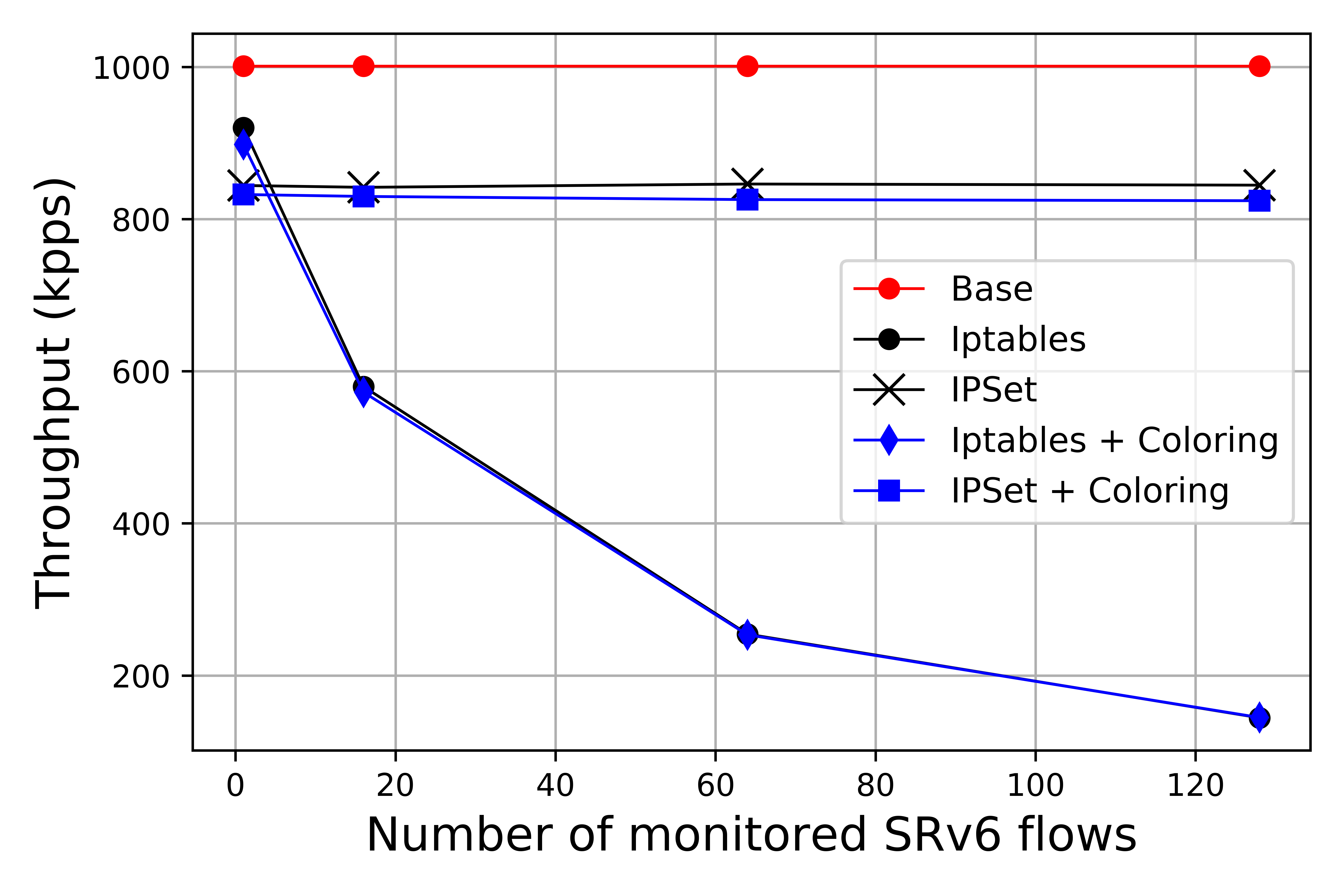}
    \caption{SUT throughput (ingress node configuration)}
    \label{fig:res1-encap}
    \vspace{-3ex}
\end{figure}

In figure \ref{fig:res1-encap}, we plot the throughput (in kpps) for the SUT configured as an ingress node considering different settings.

The red line represents the SUT base performance in the simple SRv6 case, i.e. it applies SR policies but does not perform any counting or coloring operations. In this case the measured average maximum throughput reached is about 995 kpps.

Then we assessed the performance loss due to the counting operations for both the \textit{iptables} and the \textit{IP set} solutions, varying the number of monitored flows. The SUT performs the matching operation but does not modify the packet to color it. The experiments results are reported in figure \ref{fig:res1-encap} using the black lines. As expected, using the \textit{iptables} based PF-PLM the performance degrades increasing the number of monitored SID lists (i.e. the number of \textit{iptables} rules). The measured throughput decreased in an inversely proportional way with the number of required \textit{iptables} rules. Instead, the \textit{IP set} version that uses the hashset allows to achieve a throughput that is almost constant with the number of monitored SID lists.
We note from Figure \ref{fig:res1-encap} that when we need to monitor a single flow, the \textit{iptables} based implementation achieves a higher SUT throughput with respect to the \textit{IP set} based one. In particular, the throughput degradation compared to the base case is 8\%, while the degradation for the \textit{IP set} based PF-PLM is 15.5\%. However when 16 flows are monitored the throughput of the \textit{iptables} based PF-PLM decreases by 42\%.

Finally, we evaluate the coloring cost of both solutions. In this case the degradation both in the \textit{iptables} and in the \textit{IP set} PF-PLM is less than 2.5\%. In particular in the \textit{iptables} case the maximum degradation is measured when a single rule is present and it reaches 10.3\% with respect to the base case. The coloring loss in the \textit{IP set} case is slightly less and the throughput degradation with respect to the base case is about 17.5\%.

\begin{figure}[t!]
    \centering
    \includegraphics[width=0.48\textwidth]{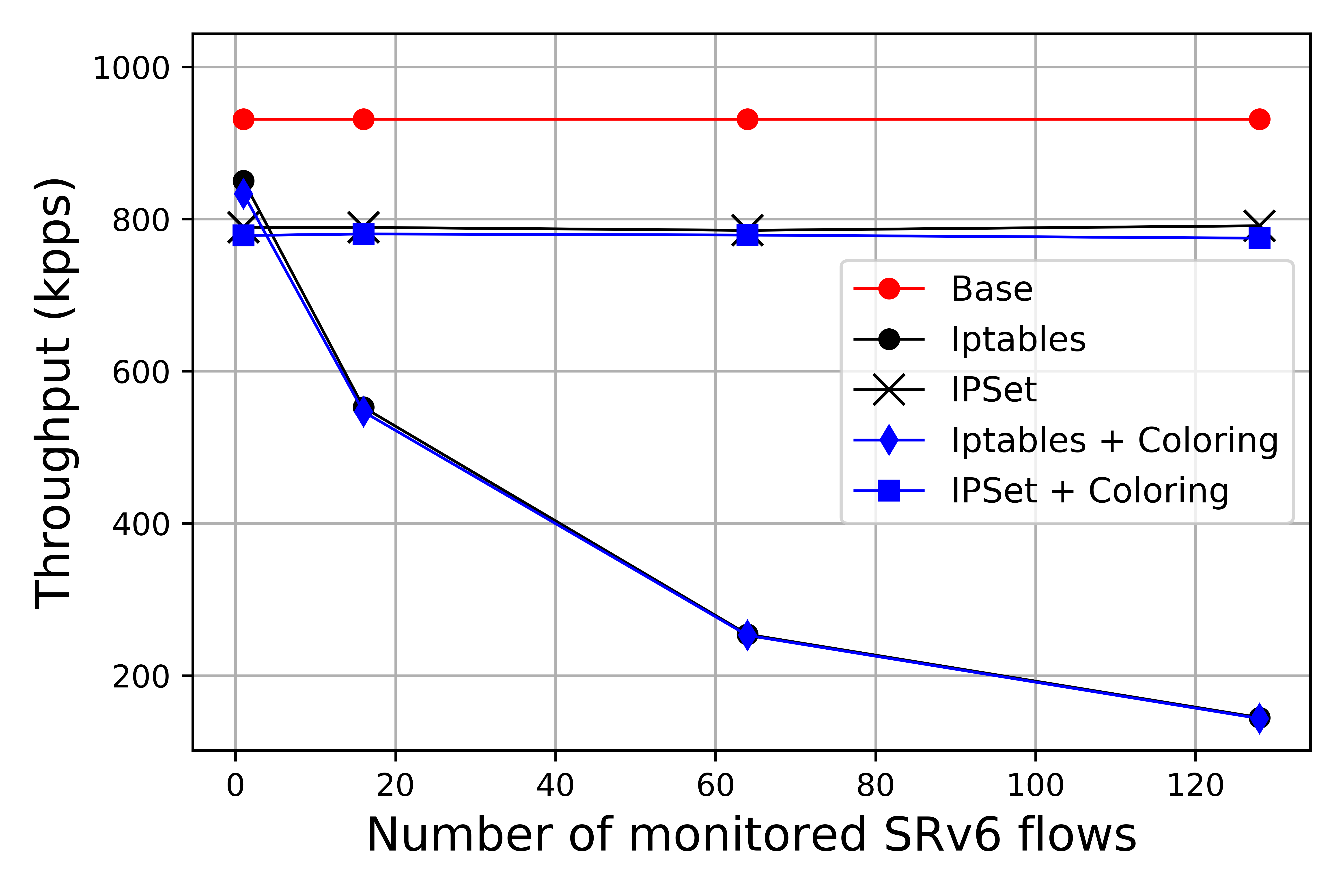}
    \caption{SUT throughput (egress node configuration)}
    \label{fig:res1-decap}
    \vspace{-3ex}
\end{figure}

In figure \ref{fig:res1-decap} we report the same analysis for the SUT configured as an egress node. We note that the decapsulation operation is more demanding and the total overall throughput in the base case is lower (about 940kpps). As for all the other configurations, the trend is similar to that of the ingress node with similar percentage degradation.

%% file: sec/9-conclusions.tex
\section{Conclusions}

In this paper we have first presented the status of the ongoing efforts for the standardization of the performance monitoring of SRv6 networks, focusing on loss monitoring. The proposed solutions consider the \textit{alternate marking} method to achieve an accurate evaluation of packet loss (single packet loss granularity). The definition of the specific packet marking mechanism for loss monitoring in SRv6 networks is left open, so we have proposed a marking mechanism based on the DS (Differentiated Services) field in the IPv6 header. 

We have provided an open source implementation of the proposed Per-Flow Packet Loss Monitoring (PF-PLM) solution in Linux. In particular, we have implemented two versions of (PF-PLM) in Linux, respectively based on plain \textit{iptables} and on \textit{IP set}.

We have setup an evaluation testbed platform, which allowed us to validate the functionality of the protocol and to evaluate some performance aspects. We were able to evaluate the cost of activating the monitoring (i.e. the degradation of the maximum forwarding throughout achievable in a node), for the two versions of the implementation. In particular, we measured the throughput degradation versus the number of monitored SRv6 flows. The \textit{iptables} based PF-PLM starts with a 8\% degradation for a single monitored flows but reaches 42\% degradation with 16 flows and 80\% degradation with 100 flows. The \textit{IP set} based PF-PLM achieves a 15\% degradation, irrespective of the number of monitored flow, hence it is scalable. We believe that 15\% degradation can be acceptable for monitoring a small subset of the flows, while it is still too high for widespread monitoring covering the large majority of the flows.

Our ongoing work includes the improvements to the counting mechanism implementation to reduce the throughput degradation. Our target is to achieve an almost negligible cost for running the loss monitoring. We are also considering the implementation of other marking mechanisms for SRv6 and the implementation of delay monitoring in addition to loss monitoring.